\documentclass[letterpaper,twocolumn,prx,aps,showpacs,floatfix,superscriptaddress]{revtex4-1}
\usepackage[latin1]{inputenc}
\usepackage{bm}
\usepackage{multirow}
\usepackage{amssymb}
\usepackage{amsbsy}
\usepackage{amsmath,  amsthm, amsfonts, mathrsfs}
\usepackage{graphicx}
\usepackage{epsfig} 
\usepackage{placeins}
\usepackage{physics}
\usepackage[usenames,dvipsnames]{color}
\usepackage[colorlinks,linkcolor=blue,citecolor=blue,urlcolor=blue,breaklinks=true]{hyperref}

\makeatletter

\begin{document} 

 \title{Stability of Quantum Dynamics under Constant Hamiltonian Perturbations}

 \author{Lars Knipschild} 
 \email{lknipschild@uos.de}
 \affiliation{Department of Physics, University of Osnabr\"uck, D-49069 Osnabr\"uck, Germany}

 \author{Jochen Gemmer}
 \email{jgemmer@uos.de}
 \affiliation{Department of Physics, University of Osnabr\"uck, D-49069 Osnabr\"uck, Germany}

\begin{abstract}
Concepts like ``typicality'' and the ``eigenstate  thermalization hypothesis'' aim at explaining the apparent equilibration of quantum systems, possibly 
after a very long time. However, these concepts are not concerned with the specific way in which this equilibrium is approached. Our point of departure 
is the (evident) observation that some forms of the approach to equilibrium, such as, e.g., exponential decay of observables,  are much more common then others.
We suggest to trace this dominance of certain decay dynamics back to a larger stability with respect to generic  Hamiltonian perturbations. 
A numerical study of a number of examples in which both, the unperturbed Hamiltonians as well as the perturbations are modelled  by partially random matrices
is presented. We furthermore develop a simple heuristic, mathematical 
scheme that describes the result of the numerical investigations remarkably well. 
According to those investigations the exponential decay indeed appears to be most stable. Dynamics that are in a certain sense at odds with the arrow of time
are found to be very unstable.
\end{abstract}


\maketitle

\section{Introduction}
While it is an empirical fact that closed (quantum) systems with many degrees of freedom 
(e.g. a solid initially prepared with some temperature-inhomogeneities, insulated from the rest of the world)
in some sense equilibrate, the question how such equilibration-processes emerge from the (reversible, non-chaotic) microscopic
principles of quantum-mechanics is still under debate. Accordingly, the eventual occurrence of equilibrium, has been investigated with great efforts, 
even and especially during the last decades \cite{gogolin_equilibration}. But recently also the specific way into the equilibrium-state \cite{gemmer_spin_relaxation},
the relaxation timescales 
of expectation values \cite{winter_equilibration_time} as well as the stability of equilibrium states \cite{farrelly_thermalization}
have gathered growing attention. A central aspect of the route to equilibrium is irreversibility. In this context,
additional to mere Loschmidt echos \cite{peres_echo}, alternative indicators have lately been employed to quantify the 
stability of the dynamics  against small, time-local perturbations \cite{larkin_otocs,kehrein_irrev}.
These indicators are based on observables, which are accessible in experiment \cite{swingle_measure_otocs,campisi_otocs}.
The typical form of the full 
relaxation dynamics of an observable the eigenstates of which are entirely unrelated to an respective Hamiltonian, has been analyzed in 
Ref. \cite{reimann_fast}, finding that this type of observable practically always relaxes much more quickly than 
many practical physical observables do. Thus, the question in which sense general principles, that appear to apply to the route to equilibrium for almost all practical
observables, emerge from the underlying quantum dynamics, remains open.

In the paper at hand, similar to Ref. \cite{reimann_fast}, we also consider observables the matrix elements of which are to some extend 
drawn at random w.r.t the eigenbasis of some  Hamiltonian. However, in order to address the large class of practically occurring, slower relaxations, 
we specifically restrict the randomness: We construct pairs of Hamiltonians and  observables $H_0, A$ such that $\langle A(t) \rangle$ conforms 
with a predefined (slow) ``relaxation function'' $g(t)$, i.e., 
\begin{equation} 
 \label{adapt}
 \langle A(t) \rangle \approx \langle A(0) \rangle g(t),
\end{equation}
for initial states to be specified below. Within the limit set by condition (\ref{adapt}), $H_0, A$ are chosen as ``random as possible'', see Sect. \ref{Model}.
This construction of $H_0, A$ serves as a basis to analyze if and how  $\langle A(t) \rangle$ changes upon the addition of a 
perturbation $V$ to $H_0$, see Sect. \ref{numerics}. 
The perturbation $V$ is also modeled by a partially 
random matrix featuring some structure to account for physically plausibility, see Sect. \ref{perturbation}. If $\langle A(t) \rangle$ does not (or only negligibly)
change due to the perturbation
we call the respective $g(t)$ stable. Note that  $\langle A(t) \rangle$ may remain stable, even though the perturbation possibly changes the 
evolution of the density matrix itself substantially (cf. Fig. \ref{fidelity}). 
Our central question is whether or not stable relaxation dynamics coincide with those observed frequently in nature. Such a coincidence 
(for which we present evidence in Sect. \ref{numerics})  hints 
in the direction of rare relaxation dynamics being rare due to their instability. In Sect. \ref{mk_model} we develop a simple heuristic scheme which describes the effect
of the perturbation on $\langle A(t) \rangle$ quite accurately. This model also supports the instability of ``unusual'' relaxation dynamics. These findings also relate 
to the concept of an ``arrow of time'': Consider a $\langle A(t) \rangle$ which is at odds with the arrow of time due to
a large recurrence, such as, e.g.  
\begin{equation} 
 \label{rec}
  \langle A(t) \rangle =\exp(-{\frac{t}{\tau}}) + \frac{1}{2}\exp(-{\frac{|t-T|}{\tau}}) \quad t > 0, T \gg \tau.
\end{equation}
Note that the example (\ref{rec}) is unrelated to a (quasi-) Poincare recurrence, since the recurrence does not appear periodically. Note furthermore that this concept of 
a violation of the arrow of time differs conceptually from the one discussed in, e.g., Ref. \cite{jennings_arrow}. \textcolor{black}{The recurrence dynamics (\ref{rec})
is
fully compatible with the quantum ``non-resonance'' equilibration principles suggested, e.g., in Refs. 
\cite{reimann_realistic,short_equilibration,lychkovskiy_conditionsEq}). It may furthermore occur for ``non-fine tuned'' initial states and is not at odds
with the eigenstate thermalization hypothesis, as will be explained in Sect. \ref{Model}. However, dynamics of this type  are hardly ever encountered in nature
Within our approach this is traced back to their instability, see Sect. \ref{arrow}}

\section{Construction of main Hamiltonian and Observable} \label{Model}
In this Sect. we describe and motivate the scheme used to construct pairs of Hamiltonians and observables $H_0, A$ in order to fulfill Eq. (\ref{adapt}). 
However, we start by 
elaborating on our choice of initial states. 

We focus on initial states which relate to the 
observables as 
\begin{equation}
\label{ini}
\rho(t=0) = \frac{\Pi_E (1 + \delta  \mathcal{A})\Pi_E }{\mathrm{Tr}(\Pi_E(1  + \delta \mathcal{A} ))}
\end{equation}
with $\delta$ sufficiently small to render $\rho$ positive. \textcolor{black}{$\Pi_E$ denotes a projector projecting onto a (possibly small) energy window of 
some Hamiltonian $ \mathcal{H}_0$, the latter is to be defined below.
 From the point of view of linear response (\ref{ini}) may be viewed as a ``response state'', state generated by the application of  a
weak static stimulus of the form $\mathcal{A}$ to an equilibrium state that essentially lives on the energy shell spanned by $\Pi_E$
\cite{kubo2012statistical,bartsch2017}. (For simplicity of notation we define $A:= \Pi_E \mathcal{A}\Pi_E$)}. Note that the choice 
(\ref{ini}) makes the evolution of the 
expectation value  $\langle A(t) \rangle$
proportional to the auto-correlation function $\mathrm{Tr}(A(t) A)$ \textcolor{black}{(Without loss of generality we choose $A$ to be traceless.) Furthermore, according to Ref. \cite{srednicki99}, expectation value dynamics 
that are simply proportional to the  respective auto-correlation function are overwhelmingly frequent in the Haar-measure invariant set of pure initial states that 
live in the above energy shell, given that the observable $A$ is  in accord with the ``eigenstate thermalization hypothesis (ETH) ansatz''. The observables considered 
below conform with the ETH ansatz by construction. Thus, expectation value dynamics which are proportional to the  respective auto-correlation function are likely 
to  result, even if the initial state is not of the form (\ref{ini}).
(We performed various non-systematic numerical checks of this statement and found it always fulfilled, see in this context also Ref. \cite{Richterprinzip_2018}.)}
Based on these considerations we now focus on tailoring the auto-correlation according to 
\begin{equation}
\label{corref}
\mathrm{Tr}(A(t) A) \approx \mathrm{Tr}(A^2) g(t)
\end{equation}

\textcolor{black}{To this end we create a $N$-dimensional Hamiltonian $H_0$ which may be viewed as the part of the above  full Hamiltonian $ \mathcal{H}_0$ which 
corresponds to the above energy shell $\Pi_E$, i.e., $ H_0=\Pi_E \mathcal{H}_0 \Pi_E$. We construct  $H_0$ by choosing its $N$ eigenvalues, 
$\{ \epsilon_j \}$, as 
uniformly i.i.d. random numbers from the interval $[-30,30]$.
 The specific form of the spectrum of this relevant part of the Hamiltonian
 reflects the 
fact that, within a sufficiently narrow energy shell, most many body Hamiltonians feature an more or less uniform density of states.} 
(A classification of Hamiltonians according to their 
level spacing  statistics (Poisson vs. Wigner-Dyson) turned out to be irrelevant here.)  
Since we are mainly interested in the thermodynamical limit (large $N$) we performed the following numerical investigations for different $N$ from 10000 to 70000.
We found all relevant quantities to converge in the limit of large $N$. The specific dimension at which this convergence is 
reached depends on the specific $g(t)$. However, we found $N=50000$ to be sufficiently large for all below discussed $g(t)$'s. 
Since diagonalizing matrices of this size 
is doable but numerically costly, we restrict our analysis to a sample of four exemplary $g(t)$, see below.

Representing an observable $A$ w.r.t. to the eigenstates of the Hamiltonian $\{\ket{\epsilon_i}\}$ yields
\begin{equation} 
 \sum_{ij} A = a_{ij} \ket{\epsilon_i}\bra{\epsilon_j}.
\end{equation}
The  auto-correlation function then reads: 
\begin{equation} \label{autocorr_A}
\mathrm{Tr}(A(t) A) = \sum\left|a_{jl}\right|^2 \cos(\omega_{jl} t) ; \quad \omega_{jl} = \epsilon_l-\epsilon_j
\end{equation}

\textcolor{black}{Since only the absolute values of $a_{jl}$ enter Eq. (\ref{autocorr_A}) and many of them correspond to the same frequencies $\omega$,
the desired dynamics (\ref{corref}) may be achieved for very many concrete choices of the $a_{jl}$. We opt for sets of matrix elements $a_{jl}$ in full agreement 
with the so called ``eigenstate thermalization hypothesis (ETH) ansatz'' for the matrix elements $a_{jl}$. The ETH ansatz reads \cite{srednicki99}:
\begin{equation} 
{a}_{jl} = {\cal A}(E) \, \delta_{jl} + 
\Omega(E)^{-1/2} \, f(E, \omega) \, R_{jl} \ ,
\label{ETH}
\end{equation}
where $E:=(E_j + E_l)/2,  \quad \omega:= E_j - E_l$. The density of states is denoted by $\Omega(E)$ and ${\cal A}(E),   f(E, \omega)$ are smooth functions of their 
arguments. Furthermore the 
$R_{jl}$ are normally  i.i.d. random real numbers  with zero mean and unit variance.
}

\textcolor{black}{
    While rigorous conditions under which the 
ETH ansatz applies are yet unknown, there are plenty of numerical examples which confirm its applicability to few body observables in non-integrable 
(in the sense of a Bethe ansatz)
many-body systems 
\cite{Rigol_ChaosETH_2016, Haque_2015, Rigol_ETH_2017, Rigol_ChaosETH_2016}. Thus employing  (\ref{ETH}) represents a valid unbiased starting
point for the investigation of dynamics of generic few-body observables in non-integrable quantum systems. For simplicity we choose : ${\cal A}(E)=0$ and $f(E, \omega)$ 
independent of $\omega$, i.e., $f=f(\omega)$. Within a sufficiently narrow energy shell $\Pi_E$, these choices are also in accord with the above cited numerical examples.
(Note that small deviations from the choice ${\cal A}(E)=0$ and their scaling with system size as discussed e.g. in 
\cite{Ueda_Atypicality_2018} 
are of no further relevance in the present context). Since in our modeling the density of states is constant by construction, i.e., $\Omega(E)=\Omega$, the above choices entail a construction of the matrix elements as 
\begin{equation}
 a_{jl} \propto f(\omega) R_{jl}
 \label{eth0}
\end{equation}
(an adequate prefactor will be determined through normalization below)}

Considering (\ref{autocorr_A}) it is evident that, in order to implement (\ref{corref}),
$f^2(\omega)$ must essentially   be chosen as  the Fourier-transform of $g(t)$. Obviously this scheme allows for the construction of almost arbitrary 
expectation value dynamics  $g(t)$ with practically arbitrary precision, given a large enough dimension $N$.


To limit numerical effort we restrict the analysis to four  different exemplary $g(t)$, see Tab. \ref{ref_functions}. 
For graphs of the functions $g(t)$ see Fig. \ref{numerical_results}.
\begin{table}
\caption{Sample of tailored expectation value dynamics $g(t)$, for graphs see Fig. \ref{numerical_results}. Despite their significant qualitative dissimilitude,
all sample dynamics are comparably ''slow``, see text.}
\centering
\begin{tabular}{ll}
\hline
\textbf{Name} & \textbf{Definition} \\
\hline 
Exponential & $g_{\mathrm{exp}}(t) = \exp(-\frac{\ln{2}}{\tau} |t|)$ \\
Oscillation & $g_{\mathrm{osc}} = \cos(\frac{2 \pi}{\tau} t) \exp(-\frac{1}{2 \tau} |t|)$ \\
Linear & $g_{\mathrm{lin}} = \begin{cases} 
	1-\frac{|t|}{2 \tau} & |t| \le 2 \tau \\
	0 & \mathrm{otherwise}
\end{cases}$ \\
Gaussian & $g_{\mathrm{gauss}}(t) = \exp(-\frac{\ln{2}}{\tau^2} t^2)$ \\
\hline
\label{ref_functions}
\end{tabular}
\end{table}
The exponential decay and the damped oscillation are chosen to represent
standard forms of equilibration dynamics which are known to occur frequently for a variety of systems. The Gaussian and especially the linear dynamics
are less common. The latter have been chosen from a larger set of "unusual" dynamics, since  - as it will turn out later - the different effects  
on  the relaxation dynamics, which are 
induced by types of perturbations, are very clearly visible for those examples. Note that all example dynamics feature  the same timescale:
They all decay to half of their initial value a time $t=\tau$, except for the damped oscillation which has its first maximum at  $t=\tau$.
We set $\tau = 15.0$ throughout this article. Note that $2\pi /\tau \approx 0.42$ is much smaller than an average frequency of the system which is on the order of 30 
 (we tacitly set $\hbar=1$ in the entire paper). In this sense all considered dynamics are slow and thus far away from the regime considered in Ref. \cite{reimann_fast}.


Numerical tests reveal that the spectra of all generated observables feature approximately semi-circular shapes, regardless of $g(t)$. 
\textcolor{black}{While semi-circular spectra may not be very  common among generic observables, the usage of random matrices featuring such
spectra has nevertheless often been a  powerful tool in the investigation of physical systems, the details of which are unknown
\cite{casati, wigner1955, wigner1957, Deutsch_ETH}.}

We normalize the operators such that the spectrum of all observables comprises eigenvalues from the interval $[-1,1]$.
Consequently, the  spectra of the different observables $A$ corresponding to different $g(t)$ are almost indistinguishable, thus they all have an almost 
identical diagonal form.
With respect to the different eigenbasises of the particular  observables, however, the Hamiltonians $H_0$ have (entirely) different eigenvectors. 
Thus the four pairs $H_0, A$ may thus be viewed as actually referring to just one single observable $A$ 
but  four different $H_0$ (with very similar spectra), giving rise to the different relaxation dynamics $g(t)$.

\section{Construction of the Perturbation} \label{perturbation}
In this Sect. the specific construction of the perturbation is described and motivated from qualitative comparison with physical models.
\textcolor{black}{Generically, a perturbation of the Hamiltonian will also consist of few-body operators. If the unperturbed Hamiltonian is 
non-integrable, modelling the perturbation according to the ETH ansatz represents a valid unbiased starting point, for the same reasons as already outlined 
below (\ref{ETH}). Furthermore there is some numerical evidence from the analysis of specific spin models confirming this approach \cite{Niemeyer_2013}.}

In order to model a generic, time independent perturbation, we thus modify the original Hamiltonian $H_0$ in the following manner:
\begin{equation}
H = H_0 + V
\end{equation}
The perturbation $V$ itself is defined regarding the eigenbasis of the observable $A$:
\begin{equation}
\label{def_perturbation}
V_{ij} = \begin{cases} 
     \sigma s(|a_j-a_i|) \mathcal{U}(-1, 1) & i \le j \\
        V_{ji} & i > j 
    \end{cases};
s(a) = \Theta(\mu - |a|)
\end{equation}
$a_j$ denote the eigenvalues of $A$ and $\mathcal{U}(-1, 1)$ 
are uniformly i.i.d. random numbers chosen from the interval $[-1, 1]$. $\Theta(x)$ is the Heaviside function.
This perturbation is more or less banded regarding the eigenbasis of $A$, the parameter $\mu \in (0, 2]$ obviously  
controls the width of the band. The value of $\sigma$ is selected such as to fulfill
\begin{equation}
\label{perturbation_strength}
\frac{\norm{V}_\textrm{HS}}{\norm{H_0}_{\textrm{HS}}} = \epsilon,
\end{equation}
($\norm{\cdots}_\textrm{HS}$ indicates the Hilbert-Schmidt norm), hence $\epsilon$ measures the strength of the perturbation. 
This parameter is fixed to $\epsilon$=0.029 throughout the entire  paper, i.e.,  we do not 
vary the strength of the perturbation, we only vary its ``bandedness'' through the parameter $\mu$.
(For the practical approximate determination of $\sigma$, see App. \ref{sigma_vs_mu}.)

 The above modeling scheme is physically motivated. Obviously at $\mu = 0$ the perturbation commutes with the observable, i.e. $[V,A]=0$. For larger  $\mu$
 the commutativity gradually vanishes, in the sense of , e.g., $\norm{[V,A]}_\textrm{HS}$ becoming larger. A range of scenarios for which stability of the 
 relaxation dynamics w.r.t. to certain perturbations is routinely expected is characterized by such a commutativity: Consider, as a first example, the (heat) energy 
 (observable $A$) in 
 some initially hot piece of 
 solid which is in thermal contact with another initially cold piece of solid, but completely insulated otherwise. Routinely, we expect the heat energy in the hot 
 solid to approach
 an equilibrium value in a certain manner. Local perturbations to the initially cold piece of solid, such as rearranging microscopic impurities, 
 replacing it (partially) by another solid featuring the same heat conductivity and capacity, slightly changing its shape without changing the interface, etc. 
 are not expected to change the way in which the heat energy of the initially hot 
 solid 
 approaches its equilibrium value. This is expected regardless of the perturbation possibly being rather significant on the microscopic level. 
 Here, the operators representing
 the above perturbations obviously commute with the local Hamiltonian of the initially hot solid, thus commutativity of perturbation and observable holds. An analogous 
 description applies to, e.g., a  spin in contact with some environment, the latter possibly being very different from the standard harmonic-oscillator bath.
 
 As a second example consider an observable which corresponds to a spatial (long wavelength)
 Fourier-component of the particle density in a many-particle lattice model. The model may comprise (short range) hopping terms, local potentials, interaction terms, etc. 
 Consider furthermore a ``local
 perturbation'', i.e., a change of the local potentials, the interactions and also the hopping terms, but such that hoppings remain short range. 
 The perturbation operator representing the change of local potentials and interactions strictly commutes with (any Fourier-component of) the particle density. 
 This does not hold for 
 the operator representing the change to the hoppings. However, also the hopping-change-operator is  banded w.r.t. to the eigenbasis of 
 the Fourier-component of the particle density. The band will be narrower for longer wavelength Fourier components and shorter hoppings. Thus, also in this example a 
 generic perturbation is banded w.r.t. to the eigenbasis of the observable.

\section{Numerical investigations of the Stability of expectation value dynamics} \label{numerics}

Before we present and  discuss results on the stability of $\langle A(t) \rangle$, we check the impact of the perturbation on the microscopic dynamics 
of the quantum state itself. To this end we adopt the quantification scheme from \cite{prosen_stability}: Consider the operator $\tilde U ^{-t} U^t$ where 
 $\tilde U^{t}$ and $U^t$ are 
the unitary time evolution operators of the perturbed and the original system, respectively. If the perturbation has very weak influence on the microscopic level,
one finds
 $\tilde U ^{-t} U^t \approx 1$. Thus $\langle \tilde U ^{-t} U^t \rangle \approx 1$ signals weak impact on the evolution of the quantum state. As 
 $0 \leq \abs{\langle \tilde U ^{-t} U^t \rangle}^2 $ is a strict lower bound , values close to zero indicate very strong influence, i.e. in this case the 
 states resulting from the perturbed and the unperturbed evolution are rather unrelated at time $t$.
 Thus we quantify the influence of the perturbation on the microscopic
 state at time $t$ by a fidelity $F$ with
\begin{equation}
F = \abs{\langle \tilde U ^{-t} U^t \rangle}^2 
\label{fid}
\end{equation}
(Note that $F$ indeed satisfies the conditions on a fidelity, see  Ref. \cite{prosen_stability})

\begin{figure}[h]
\centering
\includegraphics[width=0.50\textwidth]{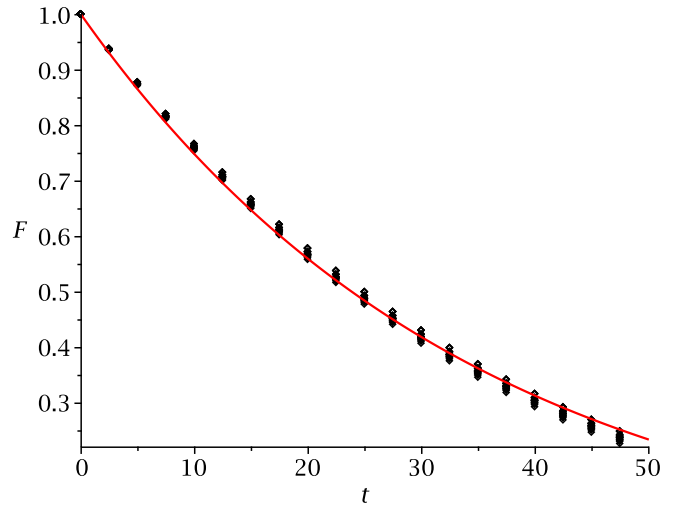}
\caption{Decay of the fidelities (\ref{fid}) for all investigated dynamics (see Tab. \ref{ref_functions}) and all perturbations, see (\ref{def_perturbation}).
For simplicity all fidelities
are represented by diamonds. The decays appear to be very similar for all considered examples. 
The red curve shows an exponential fit $F(t) =\exp(-0.029 t)$\footnote{\color{black}The coefficient in the exponent coincides with $\epsilon$. This might be traced back
to principles detailed in \cite{prosen_stability}.}. }
\label{fidelity}
\end{figure}

We display the numerical results on $F$ in Fig. \ref{fidelity}.
For all $g(t)$ and all parameters $\mu$, the fidelities decay exponentially with decay-rates of approx. 0.029
(For an analysis of the origin of the 
exponential decay, cf. Ref. \cite{prosen_stability}.) Thus, on the level of microscopic quantum states, all modeled scenarios are very similar. 
The timescale of all sample $g(t)$ is chosen to be comparable to the timescale on which the fidelity decays \footnote{The case of fidelities decaying
much faster than observables is interesting too. In this limit we found convergence to occur beyond $N=70000$ and thus did not pursue its systematic 
investigation any further. Qualitatively, however, results appear to be very similar.}. Thus, stability of 
$\langle A(t) \rangle$ cannot
be traced back to a stability of the evolution of the quantum state itself, as the latter is always strongly affected by all perturbations.

\begin{figure*}[h]
\includegraphics[width=0.49\textwidth]{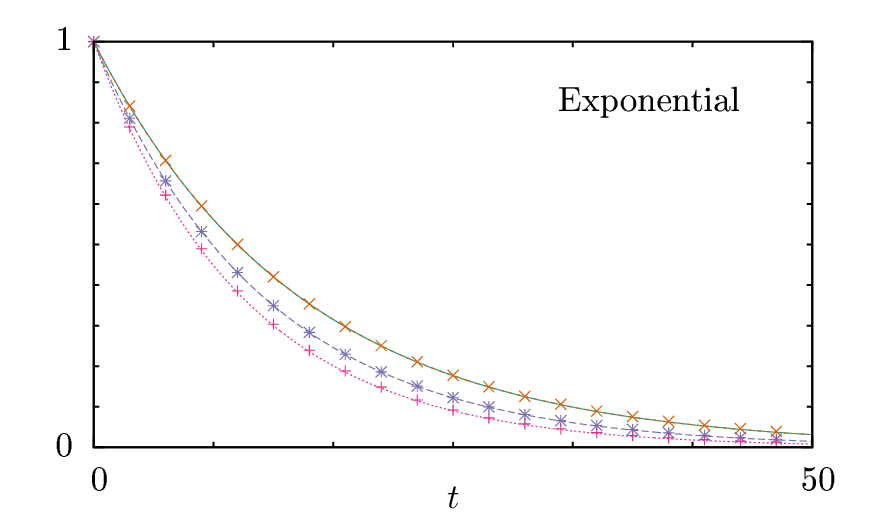}
\includegraphics[width=0.49\textwidth]{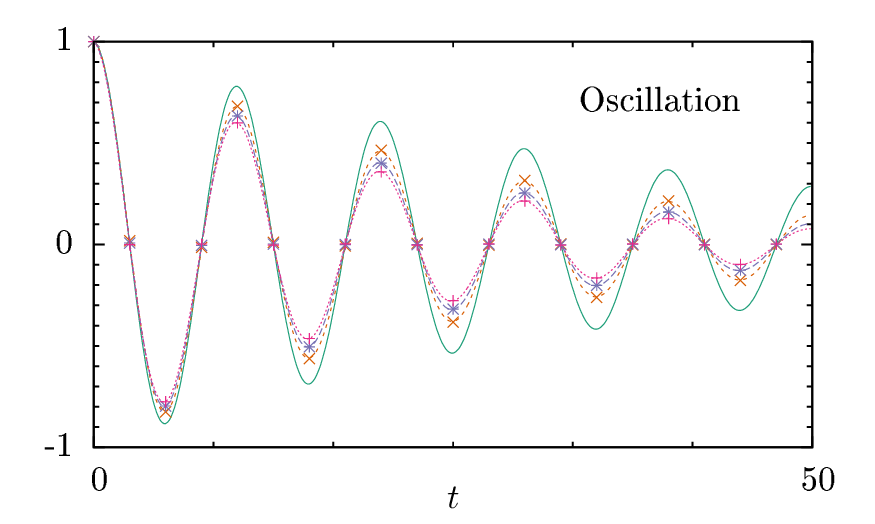}
\includegraphics[width=0.49\textwidth]{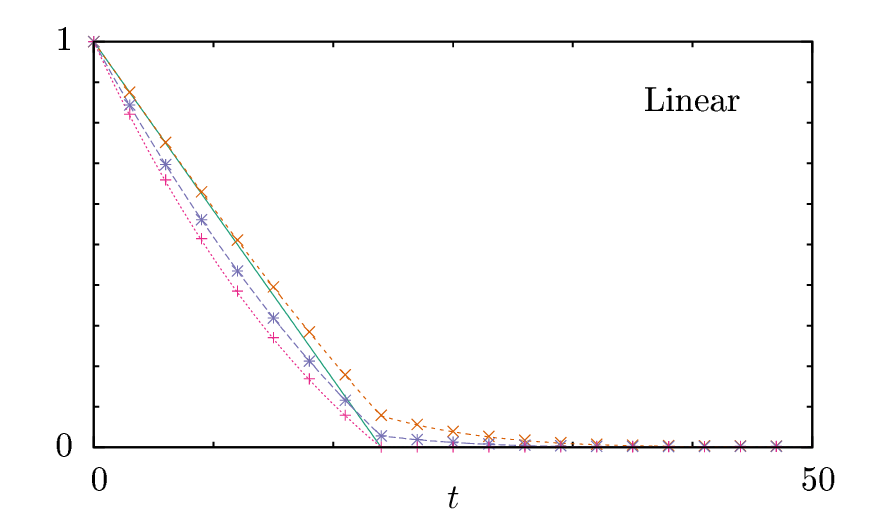}
\includegraphics[width=0.49\textwidth]{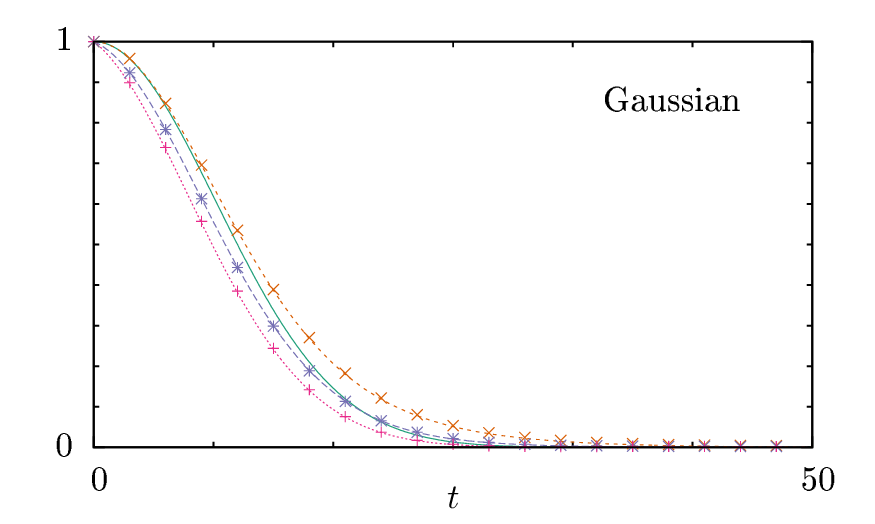}
\includegraphics{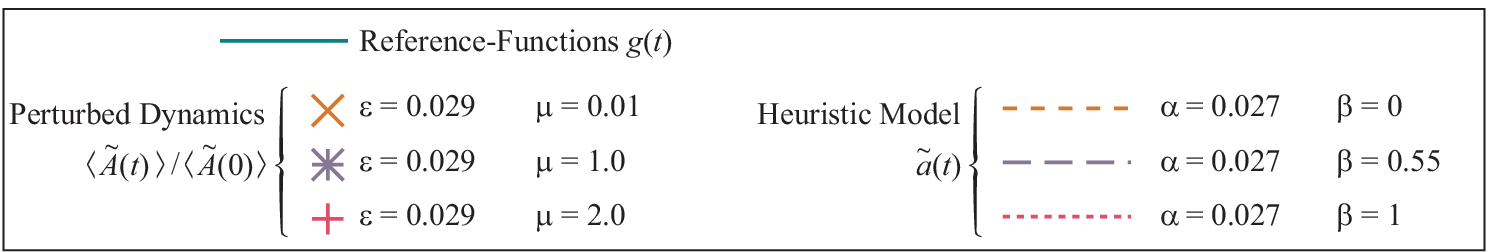}
\caption{Graphs of the reference expectation value dynamics $g(t)$ (cf. Tab. \ref{ref_functions}, solid lines), the actual quantum dynamics as
resulting from the perturbations (symbols), and the dynamics as 
generated from the  heuristic model (\ref{mk_influence}) (dashed lines). There appears to be only one case of full stability: the exponential decay remains unaltered under 
narrowly banded perturbations, even though all perturbations are equally strong. The agreement of the heuristic model with the actual quantum dynamics is very good.}
\label{numerical_results}
\end{figure*}

Now we turn to the dynamics of the expectation value of the observable $A$.
($\langle A(t) \rangle$ and $\langle \tilde{A}(t) \rangle$ denote the expectation values of the observable $A$ for the unperturbed and the
perturbated evolution, respectively.) The numerical results are displayed in Fig. \ref{numerical_results}.
{
\color{black}Obviously, in general the perturbations affect $\langle A(t) \rangle$. This influence is moderate due to perturbations being rather weak, 
but clearly visible and much larger than the statistical effects. 
}

Furthermore the effect of the perturbation does qualitatively depend on the band-width $\mu$: For example 
the relation $|\langle \tilde{A}(t) \rangle| \le |\langle A(t) \rangle|$ appears to be always valid for non-banded perturbations ($\mu=2$). \textcolor{black}{
If the principle of very slow relaxations always becoming faster upon perturbations was strictly valid, the above relation would always have to hold.}
It is, however, violated for 
banded perturbations in some cases. This may be seen nicely at the linear and the Gaussian equilibration.

Among the considered examples there is only one case in which the relaxation dynamics remains entirely unaltered, i.e., is strictly stable: 
the case of an exponential decay and a narrowly 
banded perturbation. Note that this holds for all times, also when the fidelity already has decayed substantially. This finding is our first main results. It agrees with 
the expectations formulated in the context of the examples in Sect. \ref{perturbation}.

While strict stability only applies to exponential decay and banded perturbations, exponential decay and damped oscillations may be called ``quasi-stable'' 
 under all perturbations in the following sense: exponential decay and damped oscillations are mapped onto  exponential decay and damped oscillations,
 only the parameters (such as 
 decay constants, etc. change). Such a description does neither apply to the linear nor the  Gaussian dynamics.

\section{Heuristic model of the perturbed dynamics and Memory-Kernel} \label{mk_model}

In the following we present a heuristic model, which describes perturbed dynamics of the previous Sect. rather accurately. This model is based on the notion of a memory
kernel. We define the memory-kernel $K(\tau)$, which generates some  dynamics $a(t)$, according to the following expression:
\begin{equation}
\label{mk_def}
\dv{a(t)}{t} = -\int_0^t K(t-t') a(t') \mathrm{d} t' = - K*a(t)
\end{equation}
Often such integro-differential equations of motion are used to calculate the evolution of a variable $a(t)$ from the memory kernel $K(\tau)$. But since Eq. (\ref{mk_def})
establishes a map $a(t) \Leftrightarrow K(\tau)$ it can also be used the other way round \textcolor{black}{(see Appendix \ref{calc_mk})}. To verify our below model (\ref{mk_influence}) numerically, 
we employ both directions.
Expressions like Eq. (\ref{mk_def}) appear routinely  in the context of the Nakajima-Zwanzig projection operator approach, etc.  to the expectation values of 
observables (\cite{nakajima_nz, zwanzig_nz, mori1965}).
Metaphorically speaking the memory-kernel describes the way the system remembers its history.

We proceed by describing  the original dynamics $a(t):= \langle A(t) \rangle/  \langle A(0) \rangle $ as well as the perturbed dynamics  
$\tilde{a}(t)= \langle \tilde{A}(t) \rangle/  \langle \tilde{A}(0) \rangle     $ through  integro-differential equations of motion of the type (\ref{mk_def}).
\textcolor{black}{The respective memory kernel $K(\tau)$ corresponding to the unperturbed dynamics $a(t)$ may be calculated from  $a(t)$, e.g., by means 
of a Laplace transform, see Appendix \ref{calc_mk}. From this  ``original'' memory kernel $K(\tau)$ we construct a ``perturbed'' memory kernel $\tilde{K}(\tau)$
as
\begin{equation}
\label{mk_influence}
\tilde{K}(\tau) = K(\tau) \exp(-\alpha \tau) - \beta \alpha \delta(\tau).
\end{equation}
We find that, for adequate choices of $\alpha, \beta$ the perturbed memory kernel  $\tilde{K}(\tau)$ produces almost correctly the respective perturbed dynamics 
$\tilde{a}(t)$, if inserted into a 
integro-differential equation of motion of the form (\ref{mk_def}), i.e., replacing $a(t) \rightarrow \tilde{a}(t)$ and  $K(\tau) \rightarrow     \tilde{K}(\tau)$ }.
In (\ref{mk_influence})
$\alpha$ is a non-negative
real parameter, which appears to be mainly dependent on the perturbation strength $\epsilon$. $\beta \in [0,1]$ seems to be mainly influenced
by the band width of the perturbation $\mu$. The narrow band limit  $\mu \rightarrow 0$ corresponds to $\beta = 0$, the limit of non-banded perturbations, 
$\mu \rightarrow 2$ corresponds to  $\beta = 1$. (For more details on the dependence of $\beta$ on $\mu$, see Appendix. \ref{beta_vs_mu}.) 
Equations (\ref{mk_def},\ref{mk_influence}) represent our heuristic model and our second main result. \textcolor{black}{The remarkable accuracy of this modelling may be captured from 
Fig. \ref{numerical_results}, where the dashed lines are calculated from said heuristic model.}

For the case of non-banded perturbations ($\beta = 1$) it can be shown, that Eq. (\ref{mk_influence})
is equivalent to:
\begin{equation}
\label{damping}
\tilde{a}(t) = a(t) \exp(-\alpha t)
\end{equation}
Thus, while a non-banded perturbation damps the dynamics of the observable itself, a narrow-banded perturbation damps its memory kernel.  
The stability and the quasi-stability as found and discussed in the previous Sect. \ref{numerics} are intrinsically built into the model: In general 
exponential decays and damped oscillations are transformed into exponential decays and a damped oscillation with different decay-constants, oscillation frequencies
and phaseshifts, respectively. Moreover, exponential decays are not affected at all by narrowly banded perturbations (small $\beta$).

All our numerical results are in very good agreement with the heuristic model  
(\ref{mk_influence}) as may be inferred from Fig. \ref{numerical_results}. The best  choice for the parameters $\alpha, \beta$ appears to depend only on $\epsilon, \mu$
but not (or only very weakly) on the original dynamics $g(t)$.

\section{Arrow of Time and instability of recurrence dynamics } \label{arrow}
In this Sect. we come back to the ``recurrence dynamics'' as discussed in the Introduction, see example given in  Eq. (\ref{rec}). We apply our  heuristic model 
(\ref{mk_influence}) to such recurrence dynamics $a_\mathrm{R}(t)$, thereby showing that the recurrence gets exponentially damped by both, non-banded and extremely banded 
perturbations. More concretely we find that 
\begin{equation}
\label{damped_revival}
\tilde{a}_\mathrm{R}(t) \approx a_\mathrm{R}(t) \exp(-\alpha t)
\end{equation}
holds in both cases. As $\alpha$ appears to depend on the overall perturbation strength $\epsilon$ but not on the bandedness (see text below  Eq. (\ref{rec})), 
this suggests
that banded as well as non-banded perturbations have the very same, strong damping effect on recurrences.

Justifying  Eq. (\ref{damped_revival}) for non-banded perturbations is very straight forward as it directly follows from Eq. (\ref{damping}). The case of extremely 
banded perturbations, i.e., $\mu = 0 \rightarrow \beta = 0$ is more involved. Laplace transformations may be used to find that in this case 
Eqs. (\ref{mk_influence}, \ref{mk_def})
are equivalent to 
\begin{equation}
\label{dephase_model}
\tilde{a}(t) = a(t) \exp(-\alpha t) + \alpha \int_0^t \tilde{a}(t') a(t-t') \exp(-\alpha (t-t')) \mathrm{d}t'
\end{equation}
We now focus on (scaled) expectation value dynamics $a_\mathrm{R}(t)$ that first decay within  $t \approx \tau'$ but feature one recurrence at time $T \gg \tau'$ 
which lasts 
for $2\tau'$ at most, i.e., deviates substantially from 0  only in an interval $T-\tau' \leq  t \leq T+\tau'$. Note that Eq. (\ref{rec}) conforms with this 
description, however the exponential form of decay and recurrence is not imperative, 
other forms like Gaussian, linear, etc. are also included in the present consideration. Since $a(t),  \tilde{a}(t)$ are both proportional to the respective 
auto-correlation functions, they are strictly upper bounded by their initial values $a(0) = \tilde{a}(0)=1$, i.e.,  $|a(t)| \le 1,  |\tilde{a}(t)| \le 1$. 

Thus an upper bound on the convolution-integral in Eq. (\ref{dephase_model}) may be found essentially by replacing $a_\mathrm{R}(t),  \tilde{a}_\mathrm{R}(t)$ by their maxima and restricting the 
range of integration 
to the maximum interval on which  $a_\mathrm{R}(t)$ has non-negligible values:
\begin{equation}
\abs{\int_0^t \tilde{a}_\mathrm{R}(t') a_\mathrm{R}(t-t') \exp(-\alpha (t-t')) \mathrm{d}t'} \le 3 \tau'
\end{equation}
Plugging this back into  Eq. (\ref{dephase_model}) unveils that  Eq. (\ref{damped_revival}) also holds in the case of extremely banded perturbations up to an
additive error of 
order $\alpha \tau'$. 

We sum this Sect. up as follows: Even though the above recurrence dynamics may be viewed as being at odds with the arrow of time they are in principle compatible with 
quantum mechanics. However, the recurrence peak gets damped by generic perturbations. The damping scales exponentially with both, the perturbation strength and the 
time after 
which the recurrence occurs. Hence very late recurrences (at large $T$) are very unstable against perturbation, even if the latter are weak.

\section{Conclusion and outlook}
We studied the influence of (weak) Hamiltonian perturbations on four  types of relaxation-dynamics of expectation values. 
The dynamics have been implemented in high-dimensional, partially random matrix models in accord with the eigenstate thermalization hypothesis.
We varied the degree to which the perturbations commute with the observable under consideration and analyzed the effect of this variation on the dynamics.
A heuristic model, based on integro-differential equations of motion has been demonstrated to nearly perfectly describe the numerical results.
Only exponential decay-dynamics  turned out to be stable against perturbations which approximately commute with the respective observable. However, exponential decays and 
exponentially damped oscillations get mapped onto exponential decays and 
exponentially damped oscillations by all considered sorts of perturbations. This unique stability of the exponential decay may thus be viewed as a reason for 
its very frequent
occurrence in nature. From the above heuristic model we also concluded that relaxation-dynamics that are at odds with the arrow of time are exponentially unstable.
\textcolor{black}{Besides shedding light on such rather fundamental questions of relaxation behavior, there is also a more practical merit: The heuristic model may be used to 
produce guesses on quantum dynamics under perturbation (if the unperturbed dynamics are known) at  numeric costs which are independent of the dimension of the Hilbertspace
of the respective system.}

Future research may clarify whether the principles found in the paper at hand by means of random matrix models, also apply to concrete physical models, 
such as spin systems, 
interacting particle lattice models, etc.

\acknowledgements
This work has been funded by the
Deutsche Forschungsgemeinschaft (DFG) - GE 1657/3-1.
We sincerely thank the members of the
DFG Research Unit FOR 2692 for fruitful discussions.
%
\bibliography{mybib}

\appendix

\section{Dependence of $\sigma$ on the band-size $\mu$}
\label{sigma_vs_mu}

Here we present a practical scheme for finding $\sigma$ for a given  $\mu$. The scheme is accurate to a limit set by the law of large numbers (w.r.t. the dimension $N$).
Furthermore 
we assume semi-circular spectra for the observables $A$.

Starting from Eq. (\ref{def_perturbation}), we find:
\begin{equation}
\begin{aligned}
\langle V_{ij}^2 \rangle = & \langle \mathcal{U}^2(-1,1) \rangle \sigma^2 s^2(\abs{a_i-a_j}) \\
= & \frac{1}{3} \sigma^2 s^2(\abs{a_i-a_j}).
\end{aligned}
\end{equation}
Hence the mean squared norm of $V$ may be calculated as follows:
\begin{equation}
\begin{aligned}
\langle ||V||_\mathrm{HS}^2 \rangle = & \sum \langle V_{ij}^2 \rangle \\
= & \frac{1}{3} \sigma^2 \sum_{ij} s^2(\abs{a_i-a_j}) \\
	\approx & \frac{1}{3} \sigma^2 \int_{-1}^1 \mathrm{d} a_1 \int_{-1}^1 \mathrm{d} a_2 s^2(\abs{a_1-a_2}) \eta(a_1) \eta(a_2),
\end{aligned}
\end{equation}
with  $\eta(a) = 2N/\pi \sqrt{1-a^2}$. Thus the last step is based on the assumption of semi-circular spectra.
The last step it was used that the spectra of the observables turned out to be approximately semi-circular.

\begin{figure}[h]
\caption{Dependence of $\sigma$ on the band size $\mu$}
\centering
\includegraphics[width=0.5\textwidth]{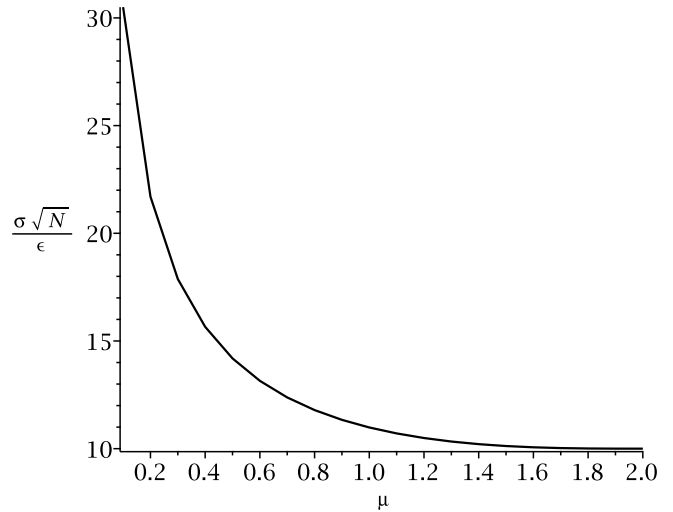}
\label{fig_sigmamu}
\end{figure}

In the same manner the $||H_0||_\mathrm{HS}$ may be approximated. 
\begin{equation}
\langle ||H_0||_\mathrm{HS}^2 \rangle = \langle \mathcal{U}^2(-30,30) \rangle N = 300 N
\end{equation}
Exploiting Eq. (\ref{perturbation_strength}) yields:
\begin{equation}
\label{sigmamu}
\frac{\sigma}{\epsilon} \sqrt{N} \approx 5 \pi \left( \int_{-1}^1 \int_{-1}^1 s^2(\abs{a_1-a_2}) \eta(a_1) \eta(a_2) \mathrm{d} a_1 \mathrm{d} a_2 \right)^{-1/2}
\end{equation}
Recall that $s(\abs{a_1-a_2})$ depends on $\mu$, (\ref{def_perturbation}). A graphic representation of Eq. (\ref{sigmamu}) is displayed in Fig. \ref{fig_sigmamu}.

\section{Determination of  parameters $\alpha$ and $\beta$}
\label{beta_vs_mu}
\begin{figure}[h]
\caption{Dependence of the parameter $\beta$ (see. (\ref{mk_influence})) on the band-size $\mu$ of the perturbation}
\centering
\includegraphics[width=0.5\textwidth]{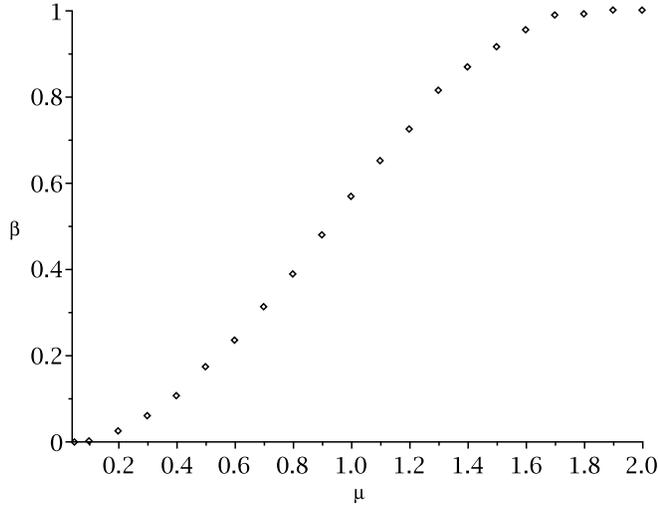}
\label{fig_betamu}
\end{figure}

As mentioned in the main text, the parameter $\alpha$ turns out to be almost independent of the band-width $\mu$ of the perturbation matrix.
It is, however, controlled by the overall perturbation strength $\epsilon$.
Accordingly, the choice $\alpha=0.027$ turns out to be appropriate for all perturbations, regardless of their bandwidth (see Fig. \ref{numerical_results}).
The numerical results indicate that $\beta$ is a function of the band-width $\mu$.
To determine  this relation, we considered the effect of varying $\mu$ on the exponential decay dynamics. 
We calculated the dependence of $\beta$ on the band-width
$\mu$ by comparing Eq. (\ref{mk_influence}) to the actual perturbed quantum dynamics and by assuming the independence of $\alpha$ on $\mu$, as mentioned above.
The result is displayed in Fig. \ref{fig_betamu}
%
%

{
\color{black}
\section{Calculating the Memory-Kernel using Laplace transforms }
\label{calc_mk}
As stated in the main text (\ref{mk_def}) establishes an implicit relation between the dynamics $a(t)$ and a memory-kernel $K(\tau)$.
This relation can be made explicit by using a Laplace transform.
By transforming (\ref{mk_def}), we find:
\begin{equation}
    \label{mk_def_laplace}
    \begin{aligned}
        s A(s) - a(0) = -\kappa(s) A(s)
    \end{aligned}
\end{equation}
$A(s)$ and $\kappa(s)$ are the Laplace transforms  of $a(t)$ and $K(\tau)$, respectively.
By algebraically transforming this equation and applying the inverse Laplace-Transformation, we calculate $K(\tau)$:
\begin{equation}
    K(\tau) = \mathcal{L}^{-1} \left\{ \frac{a_0}{A(s)} - s \right\}
\end{equation}
}

\end{document}